\documentclass[preprint,showpacs,prb,preprintnumbers,amsmath,amssymb]{revtex4}

\usepackage{graphicx}% Include figure files
\usepackage{dcolumn}% Align table columns on decimal point
\usepackage{bm}% bold math

\begin{document}
\title{Non-Fermi-liquid behavior in Ce(Ru$_{1-x}$Fe$_x$)$_2$Ge$_2$: cause and effect}

\author{W. Montfrooij$^{1)}$, M. C. Aronson$^{2)}$, B. D. Rainford$^{3)}$, J. A. Mydosh$^{4,5)}$, R. Hendrikx$^{4)}$, T. Gortenmulder$^{4,6)}$, A.P. Murani$^{7)}$, P. Haen$^{8)}$, I. Swainson$^{9)}$, and A. de Visser$^{6)}$.}

\affiliation{$^{1)}$ University of Missouri, Columbia, MO 65211, USA\\
$^{2)}$ University of Michigan, Ann Arbor, MI 48109, USA\\
$^{3)}$ University of Southampton, Southampton SO17 1BJ, United Kingdom\\
$^{4)}$ Leiden University, 2300 RA Leiden, the Netherlands\\
$^{5)}$ University of Cologne, 50937 Koeln, Germany\\
$^{6)}$ Van der Waals-Zeeman Institute, University of Amsterdam, 1018 XE Amsterdam, the Netherlands\\
$^{7)}$ Institute Laue-Langevin, BP 156, 38042, Genoble cedex 9, France\\
$^{8)}$ CRTBT, CNRS, BP 166, 38042 Grenoble cedex 9, France\\
$^{9)}$ Steacie Institute for Molecular Sciences, NRC, Chalk River, Canada KOJ 1J0}

\begin{abstract}
{We present inelastic neutron scattering measurements on the intermetallic compounds Ce(Ru$_{1-x}$Fe$_x$)$_2$Ge$_2$ ($x$=0.65, 0.76 and 0.87). These compounds represent samples in a magnetically ordered phase, at a quantum critical point and in the heavy-fermion phase, respectively. We show that at high temperatures the three compositions have the identical response of a local moment system. However, at low temperatures the spin fluctuations in the critical composition are given by non-Fermi-liquid dynamics, while the spin fluctuations in the heavy fermion system show a simple exponential decay in time. In both compositions, the lifetime of the fluctuations is determined solely by the distance to the quantum critical point. We discuss the implications of these observations regarding the possible origins of non-Fermi-liquid behavior in this system.} 
\end{abstract}
\maketitle

%\section{Introduction}
During the past decade, a multitude\cite{stewart,sachdev} of metallic systems have been uncovered in which the low-temperature response departs from that predicted by the standard Landau Fermi-liquid theory for metals. This non-Fermi-liquid (nFl) behavior is observed in systems that have been prepared to be on the brink of ordering magnetically at $T$= 0 K, the quantum critical point\cite{doniach} (QCP). Neutron scattering studies\cite{stewart,meigan2,schroeder,meigan,raymond}, probing the microscopic origins of this nFl behavior revealed two distinct types of QCP-systems. On the one hand, systems were observed\cite{raymond} in which the onset of order was given by the predictions of the spin density wave (SDW) theory\cite{hertz,millis} for itinerant electrons at a QCP, the local moments having been screened fully by the Kondo effect. On the other hand, systems were found where the QCP was given by the point where the Kondo temperature is identical zero\cite{coleman2,si,schroeder}: at this point in the phase diagram local moments first evolve and the local susceptibility diverges. However, recently it was shown\cite{montfrooij} that Ce(Ru$_{1-x}$Fe$_x$)$_2$Ge$_2$ did not fall in either category, and data on the heavy-fermion compound\cite{knafo} Ce$_{0.925}$La$_{0.075}$Ru$_2$Si$_2$ showed a scaling behavior that could not be reconciled with the above scenarios.\\

Ce(Ru$_{1-x}$Fe$_x$)$_2$Ge$_2$, prepared to be at the QCP ($x$=0.76), is a unique system\cite{montfrooij} that evolves towards an incommenserate ordered phase in a manner that cannot be described by the local moment scenario or the SDW scenario. The dynamic response of this system is characterized by a modified Lorentzian lineshape in energy and exhibits $E/T$ scaling\cite{meigan2,schroeder,meigan,montfrooij}, neither of which is predicted by the SDW-scenario\cite{hertz,millis}. In addition, the local susceptibility was found to saturate\cite{montfrooij} on approaching the QCP, while the exponents describing the dynamic and static evolution differed from each other in the temperature range 1.8 $< T <$ 20 K, neither observation being consistent with the local moment scenario\cite{si}. Instead, a change in intermoment coupling at the lowest temperatures appeared to drive the system to an ordered phase, somewhat similar to the changes in interaction strengths observed near the order-disorder phase transition in U$_2$Zn$_{17}$\cite{broholm}.\\
 
Here we report on the response of Ce(Ru$_{1-x}$Fe$_x$)$_2$Ge$_2$ when prepared to be on either side of the QCP along the chemical composition axis (see Fig. \ref{intensity}). A prelimanary account is given in Ref. \onlinecite{mmm}. We find the expected result that at high temperatures ($\sim$ 200 K) the three samples are indistinguishable. However, in contrast to the response of the QCP-sample\cite{montfrooij}, we show that at lower temperatures the decay (in time) of spin fluctuations in the heavy fermion (HF) phase is Lorentzian. At the same time, the temperature and q-dependences of the decay rates show strong similarities between the critical composition and the HF-sample.\\
 
We prepared $\sim$ 30 g polycrystalline samples of
iso-electronic Ce(Ru$_{1-x}$Fe$_{x}$)$_2$Ge$_2$ ($x$= 0.65, 0.76, 0.87) by arc-melting, followed by a two week anneal at 1000
$^{o}$C. These compositions represent an antiferromagnetically ordered compound, a compound at the quantum critical point and a compound in the heavy fermion quantum disordered part of the phase diagram\cite{fontes,montfrooij}, respectively (see inset Fig. \ref{intensity}). 
Microprobe and electron backscattering measurements verified that in each case the average composition was the intended one. Isolated regions of a
CeGe$_2$ impurity phase were found in all samples, occupying less than
2\% of the sample volume. Hence, the importance of the impurity phase is negligible for the interpretation of neutron scattering experiments (a bulk probe), but the results of resistance measurements are inconclusive.
The samples prepared to be at the QCP and in the HF-region showed good homogeneity ($x$=0.76 $\pm$ 0.02, and $x$=0.87 $\pm$ 0.02, respectively). However, the sample prepared to be in the ordered phase showed phase segregated regions with $x$ =0.59 and $x$=0.76, and consequently we only use the neutron scattering results on this sample at the highest temperatures.\\

The neutron scattering experiments were carried out at the IN6 time-of-flight (TOF) spectrometer at the Institute Laue-Langevin. IN6 was operated with incident neutron wavelength $\lambda$ of 5.12 $\AA$, yielding an 
energy resolution of 0.1 meV (Full Width at Half Maximum-FWHM). We collected data for our three samples at 11 temperatures between 1.8 K and 200 K, measuring for two hours per temperature. In addition, we measured the empty crysostat spectra at 40 K, a vanadium reference sample of the same geometry as the samples at 30 K, and an isostructural non-magnetic reference sample LaRu$_2$Ge$_2$ at 6 temperatures between 1.8 K and 200 K.\\

The corrected data show that at high temperatures the response of all three samples is governed by the paramagnetic behavior of the Cerium moments.
We corrected the sample spectra for the container and spectrometer background. This correction is small compared to the signal of the neutrons scattered by the sample. Next, the spectra were corrected for self-attenuation and absorbtion of the neutrons by the sample. In order to carry out this correction with the desired accuracy, we numerically modelled the individual buttons that make up the slab geometry, and used Sears' algorithm\cite{sears} to correct the spectra for all scattering angles and energy transfers. The spectra were then normalized to an absolute scale using the vanadium reference sample and the sample weights. Next, we corrected the data for multi-phonon scattering, based on the temperature dependence of the scattering from the isostructural LaRu$_2$Ge$_2$ sample. This correction turns out to be small, albeit non-negligible. The intensity of the incoherent scattering obtained from the fully corrected data sets agreed well with the expected incoherent intensity. We show the energy integrated intensity due to magnetic scattering for all three samples in Fig. \ref{intensity} for $T$= 200 K. This figure attests to the accuracy of the correction procedures and demonstrates that at high temperatures all three compositions have the (expected) identical response of a local moment system with an unpaired Cerium f-electron.\\

On lowering the temperature, marked differences in the magnetic response are observed amongst the three compositions (see Fig. \ref{evolution}). We focus on the critical composition and the HF-composition for the quantitative analysis. At T= 200 K all compositions show an identical response, both in wave number (see Fig. \ref{intensity}) and in energy (Fig. \ref{evolution}). In Fig. \ref{evolution} we show the imaginary part of the dynamic susceptibility $\chi$"$(q,E,T)$ divided by $E/T$ in order to obtain an energy symmetric form and to account for trivial temperature dependences. The onset of critical fluctuations is clearly observed in Fig. \ref{evolution} for the critical composition at wave number $q$= 0.4 \AA$^{-1}$ close to the magnitude of the ordering wave vecor\cite{sxtal} $|Q|$= $|$(0,0,0.45)$|$= 0.27 \AA$^{-1}$, while the total scattered intensity of HF-composition appears to show only a linear temperature dependence; however, close inspection shows that the line width for the HF-concentration decreases with temperature (see Figs. \ref{lorentzian} and \ref{linewidth}) while the total magnetic moment decreases because of Kondo screening. In fact, the static susceptibility $\chi_q(T)$ (= $\int dE\chi"/E$) of the HF-compound follows a $1/\sqrt{T}$ temperature dependence (see below).\\

In contrast to the sample prepared to be at the critical composition, we observed a Lorentzian lineshape in energy for the HF-composition. This implies that spin fluctuations decay according to a simple exponential in time. We show the best fit to a Lorentzian in Fig. \ref{lorentzian}. We note that the increased line width of the HF-composition compared to the critical composition, in combination with the reduced scattering power of the HF-compound does allow for deviations to go unnocticed at the higher temperatures, however, at the lowest temperatures the HF-composition unmistakably shows a broad (in energy) Lorentzian line shape. This in stark contrast to the lineshape of the critical fluctuations for the QCP-composition (see Fig. \ref{lorentzian}a). The amplitude of the Lorentzian lines describing the HF-compound, i.e., the static susceptibility $\chi_q(T)$, shows a $1/\sqrt{T}$  temperature dependence for 4$< T < 100 $ for all q-values (see Fig. \ref{linewidth}b), with a Kondo screening decrease at the lowest temperatures. We note that, apart from this drop, this $1/\sqrt{T}$ dependence of the static part of the response is similar to the temperature dependence of the uniform suscpetibility $\chi_0(T)$ measured\cite{montfrooij} for the critical composition.\\

The line width for the HF-composition displays a linear temperature dependence (Fig. \ref{linewidth}a), akin to the critical composition\cite{montfrooij}. Even though the T= 0 K value of the line width of the HF-composition remains finite (see Fig. \ref{linewidth}d), in both compositions the line width shows a qualitatively similar q-dependence, dropping sharply with decreasing $q$ for $q < $ 0.8 \AA$^{-1}$. Thus, on going away from the QCP, be it in temperature, wave number or composition, the line width increases with increased separation from the QCP. It is possible to express the line width as a linear combination of the distances to the QCP, with the distances $\Theta_T$, $\Theta_q$ and $\Theta_x$ measured along the $T$, $q$ and $x$-axes, respectively: $\Gamma(T) = \Theta_T$ with $\Theta_T \sim T$, $\Gamma(T,q) = a_q\Gamma(T)+\Theta_q$ with $\Theta_q$ the residual line width at the QCP shown in Fig. \ref{linewidth}d and $a_q$ practically q-independent (Fig. \ref{linewidth}a), and $\Gamma (T,q,x) = b_x\Gamma(T,q) + \Theta_x$ with $b_x$= 5 and $\Theta_x$ = 20 K (Fig. \ref{linewidth}d) for the HF concentration investigated here. At the QCP ($x=x_c$, $T$ = 0 K and $q$=$Q$) we have $\Theta_T = \Theta_q = \Theta_x$ = 0.
This strongly suggests that in both compositions the decay mechanism for spin fluctuations is similar, and demonstrates that the lifetimes of these fluctuations are determined by the distance from the QCP.\\

We argue that the above observations show that the coupling of the bosonic degrees of freedom to the fermionic ones is not of prime importance in determining the characteristics of the order-disorder phase transition. There are two ways our observations can be interpreted, both leading to the same conclusion regarding the importance of the coupling.
On the one hand, the fact that the lifetime of the magnetic fluctuations only depends on the distance to the QCP suggests that the basic underlying decay mechanism is unchanged throughout the phase diagram. The change in line shape should, in this case, be interpreted as a result of the (non-linear) coupling to the metallic medium, attributable to the screening of the local moments by the conduction electrons. Thus, this coupling produces non-Fermi-liquid behavior as a side-effect, but does not ultimately determine the characteristics of the order-disorder transition. The coupling between fermionic and bosonic degrees of freedom is not the ultimate driving force for this transition\cite{si}, the approach is determined by fluctuations in the order parameter and the overall distance to the QCP. This would also suggest that a generalized Curie-Weiss law for the magnetic fluctuations in which temperature, energy, momentum {\it and composition} are treated on equal footing should be possible for Ce(Ru$_{1-x}$Fe$_x$)$_2$Ge$_2$, but that distilling this law from experiments is hampered by the non-linear feedback mechanism.\\

On the other hand, a more exotic interpretation for this exotic system is that the non-Fermi-liquid behavior itself originates in the fact that the average lifetime of a fluctuation is determined solely by the distance to the QCP, irrespective of whether this distance is measured along the temperature axis or any other axis. For instance, a fluctuation that was caused by a local temperature variation, might decay as if it was caused by a local pressure variation, there is no intrinsic difference between the two in this QCP-system. One way of stating this is that at these very high energies (compared to the vanishing intrinsic energy scales) the system has completely lost its identity. This unusual coupling between fluctuations could well be at the heart of the non-Fermi-liquid response and the observed change in lineshape. E.g., disturbing the system with a magnetic field of the wrong wavelength, such as is done in uniform susceptibility experiments (q=0), is equivalent to disturbing the system by changing the temperature. Therefore, one should not expect the uniform susceptibility to become constant even at the lowest temperatures. As before, the coupling of the bosonic degrees of freedom to the fermionic ones does not appear to be of prime importance. Understanding the behavior near a QCP therefore equates to understanding why exactly the decay of fluctuations is determined solely by the distance to the QCP. We are currently performing single crystal experiments on Ce(Ru$_{1-x}$Fe$_x$)$_2$Ge$_2$ to scrutinize which of these two interpretations is closest to the truth.\\ 

In summary, Ce(Ru$_{1-x}$Fe$_x$)$_2$Ge$_2$ (x= 0.65, 0.76, 0.87) is a local momemt system at high temperatures, with the local moments residing on the Ce ions. With decreasing temperature, the characteristic time scale on which the moments fluctuate increases inversely proportional to the distance to the QCP, with the distance measured along the temperature, momentum and concentration axes. The decay of the fluctuations is given by an exponential in time away from the QCP (in the HF phase), however, close to the QCP, spin fluctuations no longer decay according to a simple exponential. Whether the latter reflects the coupling with the fermionic degrees of freedom, or whether it reflects coupling of bosonic fluctuations of differing origins remains an open question. Either way, it does appear that in this system the main characteristics of the order-disorder transition are determined by the fluctuations of the order parameter. Finally, the fact that the observed response is so markedly different between the two samples that were prepared using identical methods and that were of comparable homogeneity, rules out the possibility that the non-Fermi-liquid response observed\cite{montfrooij} in the sample at the QCP results from the polycrystallinity of the compound or from compositional fluctuations within the compound.\\

Work at the University of Michigan was supported by NSF-DMR-997300, work at the University of Missouri was supported by Missouri University research board grant RB-03-081.

\begin{figure*}[b]
\caption{Energy integrated magnetic intensity per formula unit (fu) for Ce(Ru$_{1-x}$Fe$_x$)$_2$Ge$_2$ over three energy ranges: $0.1 < E < 0.3$ (lower points), $0.1 < E < 0.9$ (middle), and $0.1 < E < 1.7$ (upper). The lack of a distinct q-dependence combined with the virtual identical results for all three samples show that at high temperatures Ce(Ru$_{1-x}$Fe$_x$)$_2$Ge$_2$ is a paramagnetic local moment system with the moments residing on the Cerium f-electrons. The sample compositions are indicated on the phasediagram\cite{montfrooij,fontes} in the inset. FM stands for ferromagnetic, AF for antiferromagnetic and nFl for non-Fermi-liquid behavior.} \label{intensity}
\end{figure*}
\begin{figure*}[b]
\caption{The symmetrized imaginary part of the dynamic susceptibility $\chi$"$(q,E,T)/(E/T)$ for $q$= 0.4 and 1.4 \AA$^{-1}$ for Ce(Ru$_{1-x}$Fe$_x$)$_2$Ge$_2$ ($x=0.76$ open losanges; $x=0.87$ filled circles). The various curves are offset for clarity, with zero intensity given by the horizontal lines. The response for both concentrations is identical at T = 200 K, but on cooling down the onset of critical fluctuations is clearly seen for the critical concentration, while a major part of the temperature dependence of $\chi$"$(q,E,T)$ for the HF-concentration is accounted for by the temperature dependence of the symmetrization factor $E/T$.}\label{evolution}
\end{figure*}

\begin{figure*}[b]
\caption{Lorentzian line fits to the symmetrized dynamic susceptibility for $q$= 0.5 and 1.4 \AA$^{-1}$ for the critical composition (part a), and for the HF-composition (part b). In contrast to the critical composition (only shown for $T$= 7 K), no deviations from Lorentzian line shapes are observed for the HF-compound (part b). The curves in part b are offset, with the zeroes of the intensity scale given by the horizontal lines. Part a and part b do not share the same intensity scale.} \label{lorentzian}
\end{figure*}

\begin{figure*}[b]
\caption{The temperature (a) and wave number (c) dependence of the Lorentzian line width for the HF sample. The line width follows a linear temperature dependence below 50 K, similar to the temperature dependence of the line width for the critical composition\cite{montfrooij} (solid line in (a), times 5). The static susceptibility $\chi_q(T)$ displays a $T^{-0.5}$ temperature dependence (b), with the effects of Kondo screening visible at the lowest temperatures. For comparison, we show the residual line width obtained from a fit to $\Gamma(T)$= $\Theta_q+ aT$ in part (d) (solid circles). The results for the critical composition are shown as open circles, multiplied by five. Note that both samples show a similar q-dependence, suggesting that $\Theta_q$ can be interpreted as the distance to the QCP in q-space.} \label{linewidth}
\end{figure*}
\end{document}